\documentclass[12pt,a4paper,english,superscriptaddress,aps,nofootinbib]{revtex4}
\usepackage[utf8]{inputenc}
\usepackage[T1]{fontenc}
\usepackage{amsmath,amssymb,graphicx}
\makeatletter
\usepackage{babel}
\usepackage[active]{srcltx}
\usepackage{graphicx,color}
\usepackage{changebar}
\usepackage{hyperref}
\usepackage[T1]{fontenc}
\usepackage{esint}
\usepackage{multirow}
\usepackage{dsfont}
\usepackage{ae}
\usepackage{amsmath}
\usepackage{braket}
\usepackage{mathtools}
\usepackage{slashed}
\usepackage{empheq}
\usepackage{multirow}
\usepackage{graphicx}
\usepackage{amsfonts}
\usepackage{amsmath}
\usepackage{hyperref}
\begin{document}

\title{Einstein-Bumblebee-Dilaton black hole 
 in Lifshitz spacetimes }

\author{L. A. Lessa}
\email{E-mail: leandrolessa@fisica.ufc.br}
\affiliation{Universidade Federal do Cear\'{a} (UFC), Departamento de F\'{i}sica - Campus do Pici, Fortaleza, CE, C. P. 6030, 60455-760, Brazil.}

\author{J. E. G. Silva}
\email{E-mail: euclides@fisica.ufc.br}
\affiliation{Universidade Federal do Cear\'{a} (UFC), Departamento de F\'{i}sica - Campus do Pici, Fortaleza, CE, C. P. 6030, 60455-760, Brazil.}

%\author{R. V. Maluf}
%\email{E-mail: r.v.maluf@fisica.ufc.br}
%\affiliation{Universidade Federal do Cear\'{a} (UFC), Departamento de F\'{i}sica - Campus do Pici, Fortaleza, CE, C. P. 6030, 60455-760, Brazil.}

%%%%%%%%%%%%%%%%%%%%%%%%%%%%%%%%%%%%
%\author{C. A. S. Almeida}
%\email{E-mail: carlos@fisica.ufc.br}
%\affiliation{Universidade Federal do Cear\'{a} (UFC), Departamento de F\'{i}sica - Campus do Pici, Fortaleza, CE, C. P. 6030, 60455-760, Brazil.}

%\author[ufca]{}
%\ead{r.v.maluf@fisica.ufc.br}
%%%%%%%%%%%%%%%%%%%%%%%%%%%%%%%%%%%%

%%%%%%%%%%%%%%%%%%%%%%%%%%%%%%%%%%%%
%\author[pici]{C. A. S. Almeida}
%\ead{carlos@fisica.ufc.br}

%\address[pici]{Universidade Federal do Cear\'a (UFC), Departamento de F\'isica, Campus do Pici, Fortaleza - CE, C.P. 6030, 60455-760 - Brazil}

%\address[ufca]{Universidade Federal do Cariri(UFCA), Av. Tenente Raimundo Rocha, \\ Cidade Universit\'{a}ria, Juazeiro do Norte, Cear\'{a}, CEP 63048-080, Brazil}

%%%%%%%%%%%%%%%%%%%%%%%%%%%%%%%%%%%%%

%%%%%%%%%%%%%%%%%%%%%%%%%%%%%%%%%%%%%

\begin{abstract}
We investigate the critical behavior of Lifshitz black holes in Einstein-dilaton gravity in the context of spontaneous Lorentz symmetry breaking. Considering the effects of both the bumblebee vacuum expectation value (VEV) and the fluctuations over the VEV, we obtained new asymptotically Lifshitz charged solutions in (3 + 1) dimensions. We consider the longitudinal massive mode of Lorentz Violation (LV) as thermodynamic pressure, leading us to establish an $P-V$ extended phase space. Within this framework, we derive the equation of state $P(T,V)$, and subsequently identify the critical points, which manifest as discontinuities in the specific heat at constant pressure. Following this, we compute the Gibbs free energy, revealing a first-order phase transition within the model. Finally, we determine the critical exponents, demonstrating their equivalence to those observed in the Van der Waals system.
\end{abstract}

\maketitle
\flushbottom

\section{Introduction}
Lorentz symmetry is fundamental to our understanding of matter. So far, this symmetry has been shown to be exact. Indeed, there are severe observational constraints on Lorentz-violating effects in the matter sector \cite{liberati2009lorentz}. The same is observed in the weakly coupled gravitational sector, although the constraints are weaker. However, we believe that at some energy interval, this symmetry can be violated. Some models in string theory \cite{1}, very special relativity \cite{2}, noncommutative spacetime \cite{3} and loop quantum gravity \cite{4}, among others, enable Lorentz symmetry violation in the gravitational UV regime. A framework to explore Lorentz violating theories is provided by the Standard Model Extension (SME), wherein LV coefficients lead to violation of the particle Lorentz symmetry \cite{5}. A mechanism for the local Lorentz violating is provided by a spontaneous symmetry breaking potential due to self-interacting tensor fields. The vacuum expectation value (VEV) of these tensor fields yields to background tensor fields, which by coupling to the Standard Model (SM) fields violate the particle local Lorentz symmetry \cite{6,7,8,Lessa:2020imi}. Moreover, the spontaneous Lorentz violation allows the LV terms in the Lagrangian to satisfy the Bianchi identities, a key property for the gravitational field \cite{6}.

One specific model of interest is the Kostelecký-Samuel (KS) model \cite{9}, where the field responsible for spontaneous symmetry breaking is a vector field known as the Bumblebee field. We are specifically interested in studying the KS model that includes multiple gauge fields and a negative cosmological constant, all minimally coupled to the dilaton field. This model allows us to investigate the effects of these interactions on the dynamics and properties of the system with LV. We can also identify alternative asymptotically Lifshitz black hole solutions in the literature, notably without the dilaton interaction, as referenced in \cite{sd1,sd2,sd3}. Thus, the most general action for this model, which preserves diffeomorphism invariance, is given by:
 \begin{equation} \label{action}
     S = \frac{1}{16\pi G_N}\int d^{d+1} x \sqrt{-g}\bigg[R - (\partial\phi)^2 -2 \Lambda_0 e^{-2\xi_0\phi}-\frac{1}{4} \sum_{i=1}^{N} F_{i}^{2} e^{-2\chi_i\phi} -\frac{1}{4}B^2 e^{-2\xi_2\phi} - V(B)e^{-2\xi_3\phi} \bigg].
 \end{equation}
where $G_{N}$ is the Newton gravitational constant and $\Lambda_0$ is the cosmological constant. The first term of the above action is the Einstein-Hilbert term. The scalar field $\phi$ is called the dilaton field that couples to the matter fields of the theory, where $\xi$ and $\chi$ are the coupling constants that measure this interaction. As we will come to observe, this specific field significantly alters the solutions within our problem, particularly when contrasted with the solutions obtained in the limit of $\phi=0$, see Ref.\cite{Lessa:2023yvw,Chan:1995fr}. Moreover, we have two physically different 2-forms. The first one composed of N strength fields of gauge fields is defined as $F:=F_{\mu\nu}dx^{\mu} \wedge dx^{\nu}$, where $F_{\mu\nu}=\partial_{[\mu}A_{\nu]}$. The single other 2-form defined as $B:= B_{\mu\nu}dx^{\mu} \wedge dx^{\nu}$, where $B_{\mu\nu}=\partial_{[\mu}B_{\nu]}$, defines the kinetic term of the so-called Bumblebee field $B_{\mu}$, responsible for the Lorentz Violation (LV).  Furthermore,  the term with $V$ is the potential that may depend on the $B_{\mu}$ field (self-interaction), on the metric $g_{\mu\nu}$ and even on derivatives of the field. 
The existence of a non-zero vacuum expectation value (VEV), i.e.,  $<B_{\mu}> \neq 0$ not only results in the spontaneous breakdown of Lorentz symmetry but also causes a violation of the $U(1)$ symmetry for the Bumblebee field. As demonstrated in Appendix \ref{a}, this implication results in a model possessing an additional degree of freedom when compared to an invariant vector theory under gauge transformations. In Ref.\cite{Maluf:2015hda} , the significance of this novel degree of freedom becomes evident when considering radiative corrections.

%Recently, a notable gravitational theory known as Hořava-Lifshitz (HL) gravity \cite{horava} has garnered considerable attention. Unlike a mere effective field theory, HL gravity is proposed to serve as a potential ultraviolet (UV) completion of general relativity due to its apparent power-counting renormalizability. This desirable feature is achieved by introducing higher-order spatial derivatives without accompanying higher-order time derivatives, resulting in a modification of the propagator that exhibits favorable properties. In HL gravity, this entails the introduction of a scalar field that characterizes a preferred spacelike foliation of spacetime.

 Gauge/gravity duality studies have been proving to be a very promising area in theoretical physics in recent years. 
This is mainly due to the operational ease that such a tool offers to obtain weakly coupled and computable dual descriptions of strongly coupled conformal theories \cite{maldacena1999large}. The strongly-coupled systems exhibits a scaling symmetry near critical points. When the critical fixed point is not dynamic, the more familiar scale invariance which arises
in the conformal group is given by
\begin{equation}\label{tr1}
    t \rightarrow \lambda t, \ \      x_i \rightarrow \lambda x_i
\end{equation}
where $\lambda$ is a real constant, the $t$ is time coordinate and $x_i$ are spatial coordinates. The AdS/CFT correspondence has continued to gain strength over time, with numerous works applying this duality \cite{co1,co2,co3,co4}. But we are interested in the development of the gravitational dual description of models exhibiting anisotropic scale invariance of the type
\begin{equation}\label{tr2}
      t \rightarrow \lambda^z t, \ \      x_i \rightarrow \lambda x_i
\end{equation}
where $z$ is called the dynamic exponent. The aforementioned scale symmetry is referred to as the Lifshitz scale symmetry \cite{balasubramanian2008gravity}. However, when $z=1$, the scaling becomes isotropic, which corresponds to relativistic invariance.   Hence, our specific focus is on the examination of black hole solutions generated by the LV in close proximity to the critical point characterized by a Lifshitz scale symmetry.

It has been proposed in \cite{kachru2008gravity,Mann:2009yx} the gravity duals of field theories with Lifshitz scaling should possess metric solutions that exhibit asymptotic behavior of the following form:
\begin{equation}\label{metric1}
   ds^2 = - \frac{r^{2z}}{l^{2z}}  dt^2 + \frac{l^2dr^2}{r^2} + \frac{r^2}{l^2} d\Omega^2_{d-1} .
\end{equation}
For $z=1$, which corresponds to the transformations (\ref{tr1}), we obtain the anti-de Sitter (AdS) spacetime, where the parameter $l$ represents the radius of AdS. It corresponds to relativistic invariance. On the other hand, when considering a non-trivial dynamic exponent, we refer to the metric (\ref{metric1}) as Lifshitz. There are intriguing values of $z$ to consider. For instance, at $z=2$, the theory's symmetry can be extended to the Schrödinger group. Please refer to \cite{sc1,sc2} for additional details. Moreover, it is important to note that for the Lifshitz scaling (\ref{tr2}) to hold in the metric (\ref{metric1}), the radial coordinate must be transformed as $r \rightarrow \lambda^{-1}r$.

It is important to emphasize that Lifshitz spacetime itself is not a vacuum solution to Einstein's equations \cite{taylor2008non}. Therefore, in order to obtain Lifshitz solutions, it becomes necessary to introduce matter fields. In Ref. \cite{Danielsson:2009gi}, gauge fields are employed, while in Ref. \cite{Bertoldi:2009vn}, Proca fields are utilized in the context of Lifshitz spacetime. In this context, the focus of our work is to explore Lifshitz solutions by considering the presence of gauge fields and fields that exhibit Lorentz violation. By incorporating these additional fields, we aim to investigate the interplay between gravity, gauge fields, and Lorentz violation in the context of Lifshitz spacetime. Furthermore,  as demonstrated in Ref. \cite{tarrio2011black}, the inclusion of a dilaton field gives rise to exact solutions. Significant solutions incorporating dilation, yet devoid of Lifshitz symmetry, are discussed in Refs.\cite{d1,d2}. This serves as the primary motivation for considering its presence in the action. Additionally, the inspiration for including a dilaton field arises from string theories, where in their low-energy limit, they reduce to Einstein gravity coupled with a scalar dilaton field along with other fields \cite{ortín_2004}. By incorporating the dilaton field into the gravitational action, we aim to explore the implications and potential connections between these two frameworks.

One approach to investigating the characteristics of black hole solutions is by analyzing their thermodynamics. The study of black hole thermodynamics held significant scientific importance for many years. The parallels between conventional thermodynamics and black holes are truly remarkable, encompassing various thermodynamic variables including pressure, volume, temperature, entropy, and more, as well as phase structures. The study of these phase structures is particularly crucial in the investigation of critical phenomena. Perhaps the most emblematic discovery about critical phenomena was made by Hawking and Page \cite{hawking1983thermodynamics}. They showed that there is a phase transition in the phase space of the Schwarzschild-AdS black hole. After this discovery, other phase transitions were discovered, as the first order phase transition in the charged Reissner-Nordström-AdS (RN-AdS) black hole spacetime \cite{ads}. In both of the referenced articles, the cosmological constant assumes a significant role; however, it does not contribute to the formulation of the first law of thermodynamics.

Lately, there has been a growing interest in incorporating the variability of the cosmological constant $\Lambda_0$ into the first law of black hole thermodynamics \cite{kubizvnak2012p,dayyani2018critical,Kubiznak:2016qmn}. To attain this objective, it was observed that the mass $M$ is now characterized by \textit{enthalpy} rather than internal energy. The Ref.\cite{kubizvnak2012p} demonstrated that by reevaluating the critical behavior of the AdS charged black hole, considering the cosmological constant as a thermodynamic variable, we encounter a system that closely parallels the behavior of the Van der Waals fluid. The $P-V$ space holds significant importance for the analysis of critical behavior, given its direct analogy to conventional thermodynamics. In Ref. \cite{dayyani2018critical}, we can find research focusing on the critical behavior of Lifshitz dilaton black holes through the utilization of the $P-V$ diagram. While it may initially appear unusual to consider the variation of a cosmological constant, it is justifiable in certain more fundamental theories, where certain constants arise as vacuum expectation values \cite{kubizvnak2012p}. Drawing inspiration from this, we intend to elevate the thermodynamic variable to include the massive mode of bumblebee fluctuations, which precisely arise from a dynamic process. Thus, we are suggesting here is to conduct an examination of the critical behavior within the expanded $P-V$ space of a Lifshitz black hole, influenced by the spontaneous breaking of Lorentz symmetry. Ultimately, our findings reveal a system with critical exponents identical to those of the Van der Waals fluid.

Before delving into the examination of Lifshitz black hole solutions, it is worthwhile to briefly explore a potential link between the Ho\v{r}ava-Lifshitz theory and models featuring spontaneous Lorentz symmetry breaking, particularly the Tensor-Vector theories. Taking inspiration from anisotropic scale invariance (\ref{tr2}), Ho\v{r}ava puts forward the concept of a renormalizable UV completion of general relativity, which does not adhere to Lorentz invariance \cite{PhysRevD.79.084008}. 
The Lorentz symmetry is violated in UV since time and space are treated differently. On the other hand, this symmetry can be violated locally through a spontaneous symmetry breaking \cite{1}. However, the Ho\v{r}ava theory presents problems such as instabilities that arise precisely due to the presence of a nondynamical spatial foliation in the action. In an attempt to address this issue, the Ref.\cite{PhysRevD.81.101502} constrained the Lorentz-violating vector field, known as the aether (a dynamic timelike vector), to be hypersurface orthogonal. Consequently, it has been shown that by doing so, the theory is identical to the IR limit of the extension of Ho\v{r}ava gravity\cite{PhysRevLett.104.181302}. 

This paper is organised as follows. In section \ref{2}, we obtain the black hole solution for the model with bumblebee fluctuations and study its geometric properties. In section \ref{3}, we study the thermodynamics of the solution found and then study the critical phenomena associated with the model. The paper concludes in section \ref{con}. We will be using units where the speed of light, Planck’s constant, and Boltzmann’s constant equals unity, $c$ = $\hbar$ = $k$ = 1. We shall also take the Lorentzian signature for the spacetime metric to be $(-, +, . . . , +)$.

%In this work, we are interested in gravitational solutions that are asymptotically (\ref{metric1}). The metric of gravitational theories more general that provide a gravity description of Lifshitz fixed points can be described as {\color{red}{NECESSÁRIO MOTIVAR ESSA GENERALIZAÇÃO}}
%\begin{equation}\label{metricgeral}
%    ds^2 =  - \frac{r^{2z}}{l^{2z}} f(r) dt^2 + \frac{l^2dr^2}{r^2f(r)} + r^2 d\Omega^2_{d-1} 
%\end{equation}
% where $f(r)$ is the blackening function. Moreover, we have that this function has the following condition
% \begin{equation}\label{assin}
%     Lim_{r \rightarrow \infty} f(r) = 1,
% \end{equation}
%so that the metric is asymptotically given by (\ref{metric1}).

%We also have $N$ 2-form set to $F:=F_{\mu\nu}dx^{\mu} \wedge dx^{\nu}$ and $B:= B_{\mu\nu}dx^{\mu} \wedge dx^{\nu}$, where $F_{\mu\nu}=\partial_{[\mu}A_{\nu]}$  and $B_{\mu\nu}=\partial_{[\mu}B_{\nu]}$. Furthermore,  the term with $V$ is the potential that may depend on the $B_{\mu}$ field (self-interaction), on the metric $g_{\mu\nu}$ and even on derivatives of the field. 

%Note that due to the $V$ potential, the $B_{\mu}$ vector field violates the $U(1)$ symmetry, so only the $F_2$ field is a gauge field. Thus, the (\ref{action}) action is composed of a gauge field, a cosmological constant, a dilaton field and a self-interacting field. All fields coupled to the dilaton field, whose coupling strength is measured by the respective coupling constants $\xi_i$.

\section{Bumblebee Excitations black hole in Lifshitz gravity}\label{2}
In this section, our aim is to seek (3+1)-dimensional black hole solutions characterized by bumblebee excitations that exhibit asymptotically Lifshitz behavior. To this end, we consider the following line element as our starting point \cite{za}:
\begin{equation}\label{metricgeral}
ds^2 =  - \frac{r^{2z}}{l^{2z}} f(r) dt^2 + \frac{l^2dr^2}{r^2f(r)} + r^2 d\Omega^2_{2} 
\end{equation}
 where the $d\Omega^2_{2}$ the metric of a unit-radius $S^2$  and the $f(r)$ is the blackening function. Thus, we impose the condition that the function $f(r)$ satisfies the following requirement:
\begin{equation}\label{assin}
\lim_{r \rightarrow \infty} f(r) = 1,
\end{equation}
ensuring that the metric asymptotically approaches the form given by (\ref{metric1}). Additionally, we specifically consider the static and symmetrically spherical scenario, where all fields vary exclusively along the radial direction of the asymptotically Lifshitz spacetime.

Taking inspiration from the work \cite{kos1}, we consider the sector of Lorentz violation to be governed by bumblebee excitations in the linear regime of fluctuations. By adopting a quadratic and smooth potential, a massive mode can be obtained through a mechanism known as the alternative Higgs mechanism. In addition to the massive mode, there exists a massless mode corresponding to Nambu-Goldstone bosons. Considering the metric (\ref{metricgeral}) that exhibits both spherical and temporal symmetries, we further assume that the vacuum expectation value (VEV) has only one non-zero radial component, see Appendix \ref{a}. Under these conditions, the following action can be derived: 
 \begin{equation} \label{action1}
     S = \frac{1}{16\pi G_N}\int d^{4} x \sqrt{-g}\bigg[R - (\partial\phi)^2 -2 \Lambda_0 e^{-2\xi_0\phi}-\frac{1}{4} \sum_{i=1}^{N} F_{i}^{2} e^{-2\chi_i\phi} -\frac{1}{4}\Tilde{F}_{2}^2 e^{-2\xi_2\phi} - V_0 e^{-2\xi_3\phi} \bigg].
 \end{equation}
Here, we define $\tilde{F} = \tilde{F}_{\mu\nu} dx^{\mu} \wedge dx^{\nu}$, where $\tilde{F}_{\mu\nu} = \partial_{[\mu} \tilde{A}_{\nu]}$ with the condition $\tilde{A}^r = 0$ when $b_r \neq 0$. Additionally, we introduce the notation:
\begin{equation}
V_0 \equiv 2 \lambda b^2 \beta_0^2.
\end{equation}
In practice, the model consists of $N+1$ vector fields combined with two Liouville potentials for the dilaton. However, it is important to note that, as we will later demonstrate, the $N$ gauge fields primarily serve as auxiliary fields within the framework

With the given action at our disposal, we can derive the equations of motion (EoM) by varying the action (\ref{metric1}) with respect to the metric, the vector fields, and the scalar field. This leads us to the following equations, respectively:
\begin{equation}\label{eq1}
    R_{\mu\nu} = \Lambda_0 e^{-2\xi_0\phi}g_{\mu\nu} + \sum_{i=1}^{N}\frac{e^{-2\chi_i \phi}}{2}T^{EM,i}_{\mu\nu}+\frac{e^{-2\xi_2\phi}}{2}T^{BUM}_{\mu\nu} +\frac{V_0}{2} e^{-2\xi_3\phi}g_{\mu\nu} + \frac{1}{2}\partial_{\mu}\phi \partial_{\nu}\phi,
\end{equation}
\begin{equation}\label{eq2}
    \sum_{i=1}^{N}D_{\mu}(e^{-2\chi_i \phi}F^{\mu\nu}_i)=0,
\end{equation}
\begin{equation}\label{eq3}
    D_{\mu}(e^{-2\xi_2 \phi}\Tilde{F}^{\mu\nu})=0,
\end{equation}
\begin{equation}\label{eq4}
    \Box \phi + 2 \xi_0 \Lambda_0 e^{-2\xi_0\phi} + \sum_{i=1}^{N}\frac{\chi_i}{2} F_{i}^2 e^{-2\chi_i\phi} + \frac{\xi_2}{2} \Tilde{F}_2^2 e^{-2\xi_2\phi} + \xi_3 V_0 e^{-2\xi_3\phi}.
\end{equation}
where 
\begin{equation}\label{eq5}
    T^{EM,i}_{\mu\nu} = (F_i)_{\mu} \ ^{\sigma}(F_i)_{\nu\sigma}-\frac{1}{4}g_{\mu\nu}F^2_i
\end{equation}
\begin{equation}\label{eq6}
    T^{BUM}_{\mu\nu} = \Tilde{F}_{\mu} \ ^{\sigma}\Tilde{F}_{\nu\sigma}-\frac{1}{4}g_{\mu\nu}\Tilde{F}^2_2.
\end{equation}
Note that we used in Eq. (\ref{eq1}) the fact that the two quantities above have traceless in 4-dimensions .

Assuming that $(F_i)_{rt}\neq 0$ and $\tilde{F}_{rt}\neq 0$, indicating an electric field configuration, we can derive the following expressions from Eqs. (\ref{eq2}) and (\ref{eq3}), respectively, for the ansatz (\ref{metricgeral}):
\begin{equation}\label{camp1}
(F_i)_{rt} = \frac{q_i e^{2\chi_i \phi}}{r^{3-z}},
\end{equation}
\begin{equation}\label{camp2}
\tilde{F}_{rt} = \frac{\tilde{q} e^{2\xi_2 \phi}}{r^{3-z}},
\end{equation}
where $q_i$ and $\tilde{q}$ are integration constants. These constants are related to the total charge through further considerations
\begin{equation}\label{carga1}
    Q_i = \frac{1}{16\pi G_N} \int e^{-2\chi_i\phi}\star F_i,
\end{equation}
\begin{equation}\label{carga2}
    \Tilde{Q} = \frac{1}{16\pi G_N} \int e^{-2\xi_2\phi}\star \Tilde{F}.
\end{equation}
This last charge is related to the transverse mode that originated through a spontaneous Lorentz symmetry breaking.

Substituting the solutions (\ref{camp1}) and (\ref{camp2}) in the components $tt$ and $rr$ of Eq.(\ref{eq1}), we can find from the combination $R^{t}\ _{t}-R^{r}\ _{r}$ that
\begin{equation} \label{dilaton}
    \phi = \phi_0 + \phi_1 \text{ln} r,
\end{equation}
where $\phi_0 $ is a integration constant and $\phi_1$ is given by
\begin{equation}
    \phi_1 =  2\sqrt{z-1}.
\end{equation}
From the expression for $\phi_1$ presented above, it is evident that we require $z\geq1$. It is worth noting that besides the context of dilaton and Lifshitz black hole solutions, scalar fields of the form (\ref{dilaton}) are also solutions in various models of asymptotically Anti-de Sitter (AdS) spacetimes \cite{dilaton1, dilaton2}. To obtain the function $f(r)$, we can substitute Eqs. (\ref{camp1}), (\ref{camp2}), and (\ref{dilaton}) into the Einstein equations (\ref{eq1}). By performing this substitution, we obtain the following expression for $f(r)$
\begin{align}\label{sol1} \nonumber
    &f(r) = \frac{l^2}{r^2 z^2}-m r^{-z-2} -\frac{\Lambda_0  l^2 e^{-2 \xi_0 \phi_0} (\xi_0 \phi_1+1) r^{-2 \xi_0 \phi_1}}{z (-2 \xi_0 \phi_1+z+2)} -\frac{l^2 V_0 e^{-2 \xi_3 \phi_0} (\xi_3 \phi_1+1) r^{-2 \xi_3 \phi_1}}{2 z (-2 \xi_3 \phi_1+z+2)} \\
    & + \sum_{i=1}^{N}\frac{q^2_i l^{2 z} e^{2 \chi_i \phi_0} (-\chi_i \phi_1+2 z-5) r^{2 \chi_i \phi_1-4 z+8}}{4 z (3 z-2 (\chi_i \phi_1+5))}+\frac{\Tilde{q}^2 l^{2 z} e^{2 \xi_2 \phi_0} (-\xi_2 \phi_1+2 z-5) r^{2 \xi_2 \phi_1-4 z+8}}{4 z (3 z-2 (\xi_2 \phi_1+5))}.
\end{align}
In the expression above, the integration constant $m$ is related to the mass, as we will discuss later. To obtain a solution that is asymptotically Lifshitz, it is crucial for the condition (\ref{assin}) to hold. In order to achieve this, we fix the cosmological constant term. Consequently, from Eq. (\ref{sol1}), we need to determine the coupling constant $\xi_0$ as follows
\begin{equation}\label{c1}
    \xi_0 = 0.
\end{equation}
That is, by making this particular choice, the dilaton field does not couple to the cosmological constant term, resulting in a pure cosmological constant contribution. Certainly, we could apply a similar approach to the other terms in Eq. (\ref{sol1}), such as the term with $V_0$. However, choosing a specific value for $V_0$ would fix the contribution from Lorentz violation (LV), resulting in a solution that does not exhibit the effects of LV. Since the main objective of the article is to investigate the effects of LV, it is not desirable to fix the term associated with LV in this manner.

It is important to note that we have not yet utilized the equation of motion (EoM) of the scalar field. This equation will yield an algebraic equation that assists in relating the parameters of the theory. By substituting the solution (\ref{sol1}) into Eq. (\ref{eq4}), we obtain the following expression:
\begin{align}\label{alg}
   &  - \Lambda_0 \phi_1 + \frac{V_0[4z\xi_3-\phi_1(1+\xi_3\phi_1)]e^{-2\xi_3 \phi}}{2} +\frac{\phi_1}{r^2} = - \frac{1}{4} \bigg[ \sum^{N}_{i=1} F_{i}^2 [4\chi_i + \phi_1]e^{-2\chi_i\phi}  + \Tilde{F}^2 [4\xi_2 + \phi_1]e^{-2\xi_2\phi}\bigg]
\end{align} 
This equation can be solved through various ways. However, our objective is to obtain a solution that highlights not only the contribution from the LV charge $\tilde{q}$ but also the effects of the LV massive mode $V_0$ discussed in the paper. To achieve this, we can choose the first gauge field to cancel the cosmological constant term and the second gauge field to cancel the third term on the left side in the Eq.(\ref{alg}). By making this choice, we find the following expression:
\begin{equation}\label{c2}
    \chi_1 = \frac{2}{\phi_1}, \ \ \chi_2 = \frac{1}{\phi_1}. 
\end{equation}
So that the first is fixed by the cosmological constant through
\begin{equation}\label{q1}
   q^2_1 = \frac{4   (1-z) l^{2-2 z} e^{-\frac{4 \phi_0}{\phi_1}}\Lambda_0}{z+1}
\end{equation}
and we use the second charges $U(1)$ to cancel the term $\frac{\phi_1}{ r^2}$ with
\begin{equation}\label{q2}
    q^2_2 = \frac{4 (z-1) l^{2-2 z} e^{-\frac{2 \phi_0}{\phi_1}}}{z}.
\end{equation}
Furthermore, the remaining $N-2$ gauge field can be canceled for the following choice
\begin{equation}
    \chi_j = -\frac{\phi_1}{4},
\end{equation}
where $j=2, ..., N-1$.  Finally, in order to cancel the transverse and longitudinal mode terms of the LV in Eq. (\ref{alg}), respectively, we have that
\begin{equation}
    \xi_2 = -\frac{\phi_1}{4} , \ \  \xi_3 = \frac{\phi_1}{4}.
\end{equation}

As our primary interest lies in exploring the contributions associated with the spontaneous breaking of Lorentz symmetry, we will henceforth consider only two $U(1)$ fields. Consequently, these fields will not contribute to the metric, illustrating their auxiliary nature in this particular configuration, since they are already fixed in (\ref{q1}) and (\ref{q2}). It is important to note that in order to obtain a black hole solution using the action (\ref{action1}), it is necessary to incorporate at least three vector fields, as previously mentioned in reference \cite{tarrio2011black}. Therefore, in this case, only two gauge fields are sufficient to achieve a black hole solution with spherical symmetry. This is because the role of the additional vector field required for spherical symmetry will be played by the transverse mode of the Lorentz violation. This rationale justifies why we no longer utilize the remaining $N-2$ gauge fields in our analysis. Moreover, as assumed in (\ref{c1}), we consider the cosmological constant term to be unity, which imposes the condition that
\begin{equation}
\Lambda_0 = -\frac{(z+1) (z+2)}{2 l^2}.
\end{equation}%Note that at $z=1$, we recover the usual result: $\Lambda_0=-\frac{3}{l^2}$.

Considering all of the aforementioned considerations, we can conclude that the solution is characterized by only four free parameters: $m$, $\tilde{q}$, $V_0$ and $\phi_0$.
 %Finally, we find that the function $f$ is given by
%\begin{align}
 %   f(r)=1- \frac{m}{r^{z+2}}  +\frac{l^2}{r^2 z^2} + \frac{V_0 l^2 e^{-\phi_0\sqrt{z-1}}}{2(z-4)r^{2z-2}}+\frac{\Tilde{q}^2 l^{2 z}  e^{-\sqrt{z-1} \phi_0}}{4 z r^{2 (z+1)}}
%\end{align}
%{\color{red}{HAIR?}}
Summarizing, the solution we found to action (\ref{action1}) is
\begin{equation}\label{resumo1}
    ds^2 =  - \frac{r^{2z}}{l^{2z}} f(r) dt^2 + \frac{l^2dr^2}{r^2f(r)} + r^2 d\Omega^2_{d-1} 
\end{equation}
where
\begin{align}\label{resumo2}
    f(r)=1- \frac{m}{r^{z+2}}  +\frac{l^2}{r^2 z^2} + \frac{V_0 l^2 e^{-\phi_0\sqrt{z-1}}}{2(z-4)r^{2z-2}}+\frac{\Tilde{q}^2 l^{2 z}  e^{-\sqrt{z-1} \phi_0}}{4 z r^{2 (z+1)}}.
\end{align}
\begin{equation}\label{resumo3}
    (F_1)_{rt} = 2\sqrt{\frac{z-1}{z}}\frac{l^{1-z}e^{\frac{\phi_0}{2\sqrt{z-1}}}}{r^{1-z} },
\end{equation}
\begin{equation}\label{resumo4}
    (F_2)_{rt} =   \sqrt{2(z-1)(z+2)}l^{-1}e^{\frac{\phi_0}{\sqrt{z-1}}}r^{z+1}, 
\end{equation}
\begin{equation}\label{resumo}
    \tilde{F}_{rt} = \frac{\Tilde{q}e^{-\sqrt{z-1}\phi_0}}{r^{z+1}},
\end{equation}
\begin{equation} \label{resumo5}
    \phi = \phi_0 + 2\sqrt{z-1} ln r, 
\end{equation}
where $\Lambda_0 = -\frac{(z+1) (z+2)}{2 l^2}$ and $V_0 = 2 \lambda b^2 \beta_0^2$. Note that a special limit for the $z$ parameter is when $z=1$. Firstly, the scalar field becomes constant for this limit. Further, the fixed-charges $q_{1,2}$ vanish and the cosmological constant is given by $\Lambda_0=- \frac{3}{l^2}$. Therefore, we recover the charged black hole solution, i.e.,  
\begin{align} \label{sol1}
    ds^2 = -\bigg( 1 - \frac{m}{r} + \frac{\Tilde{q}^2}{r^2} -  \Lambda_{eff}r^2 \bigg) dt^2 + \bigg(  1 - \frac{m}{r} + \frac{\Tilde{q}^2}{r^2} -  \Lambda_{eff}r^2  \bigg) ^{-1}dr^2 + r^2 d\Omega^2_{2}.
\end{align}
where the effective cosmological constant is given by 
\begin{equation}
    \Lambda_{eff} \equiv \Lambda_0 + \lambda b^2 \beta_0^2.
\end{equation}
Result similar to that found in \cite{leandro}. Thus, we can recover the AdS charged black holes \cite{ads} for $\lambda b^2 \beta_0^2<\frac{3}{l^2}$. On the other hand, for $z\neq1$, the solution exhibits significant differences compared to the AdS charged black holes. This highlights the profound impact of the dilaton field on the solution, showcasing the drastic changes it induces. It is precisely this dramatic change that we will further investigate when addressing the critical behaviors of the solution (\ref{resumo1}).

Now, let us analyze some geometric properties of the solution (\ref{resumo1}). First, we can determine the position of the horizon $r_h$ by setting $f=0$. This results in the following algebraic equation for $r_h$:
\begin{equation}\label{hor}
    r_h^{2z-2} - m r_h^{z-4} +\frac{l^2}{z^2}r_h^{2z-4} + \frac{ V_0 l^2 e^{-\sqrt{z-1}\phi_0}}{2(z-4)} + \frac{\Tilde{q}^2 l^{2z}e^{-\sqrt{z-1}\phi_0}r_h^{-4}}{4z}=0,
\end{equation}
In this case, we assume that $r_h$ represents the largest positive real root of the equation $f=0$. Unfortunately, an exact solution for $r_h=r_h(m,\tilde{q}, \phi_0, V_0)$ is not readily available. However, as noted by Ref.\cite{tarrio2011black}, the mass parameter $m$ is not a fundamental parameter of the theory. Instead, the fundamental parameters are $\tilde{q}$, $\phi_0$, $V_0$ and the temperature $T$. Motivated by this, we can solve Eq. (\ref{hor}) with respect to the mass, yielding
\begin{equation} \label{massa}
   m = r_h^{z+2} \left(\frac{\Tilde{q}^2 l^{2 z} r_h^{-2 (z+1)} e^{-\sqrt{z-1} \phi_0}}{4 z}+\frac{ V_0 l^2  r_h^{2-2 z} e^{-\sqrt{z-1} \phi_0}}{2 (z-4)}+\frac{l^2}{r_h^2 z^2}+1\right).
\end{equation}
This quantity will play a crucial role as we utilize it to derive an expression for the temperature that depends solely on $\tilde{q}$, $r_h$, $\phi_0$ and $V_0$. Additionally, as we will explore later, in the limit when the black hole becomes extremal, i.e., $T=0$, the near-horizon geometry is given by $AdS_{2} \times S^2$.

As mentioned, certain studies suggest that the cosmological constant can exert a form of pressure on black holes, and this consideration is crucial for the investigation of criticality in systems. Drawing inspiration from these works and recognizing the similarities between the cosmological constant and the LV massive mode, it is reasonable to speculate that this mode may also exert pressure on the black hole. From Eq. (\ref{eq1}), it is possible to associate a pressure with $V_0$. Consequently, it can be shown that the pressure due to the massive mode is given by\footnote{We can model the stress-energy tensor as $T_{\mu\nu} = \text{diag}(\rho, -P_r, -P_t, -P_t)$, where $\rho$ is the energy density and $P_r$, $P_t$ are the radial and tangential pressures, respectively. By utilizing Eq. (\ref{eq1}), we can determine the pressures associated with the parameters $V_0$ and $\Lambda$. Considering that Lifshitz spacetime is non-isotropic, we account for a non-isotropic fluid. Furthermore, we focus solely on the radial component since the tangential components do not contribute to the work exerted by the pressure on the black hole.}
%We can see this by assuming that the black hole described by solutions (\ref{resumo1})-(\ref{resumo5}) can be modeled by stress-energy tensor $T^{\mu} \ _{\nu} = \frac{1}{8 \pi G_N} (-\rho, P_r, P_{t1}, P_{t2})$, where $\rho$ is the energy density, $P_{r}$ is the radial pressure and $P_{t1,2}$ are the tangent pressure.  Clearly, we do not have a fluid that describes the black hole with isotropic behavior, since Lifshitz spacetime (\ref{metric1}) naturally introduces an anisotropy. Furthermore, it is natural to expect the two tangential pressures to be identical, since spherical symmetry is preserved in this context. From Einstein equations with the solution (\ref{resumo1}), we get that
\begin{equation}\label{pr}
    P_{LV} = \frac{-V_0 e^{-\sqrt{z-1} \phi_0}}{16 \pi  G_N r_h^{2z-2}}.
\end{equation}
Note that due to the fact that $V_0>0$, the pressure obtained above is negative, bearing resemblance to the pressure associated with a de Sitter solution. On the other hand, the pressure associated with the cosmological constant can be expressed as:
\begin{equation}\label{presslam}
    P_{\Lambda_0} = \frac{z+2}{8 \pi  G_N l^2}.
\end{equation}
Note that this pressure is positive and in the limit $z=1$ we recover the usual case $ P_{\Lambda_0} = \frac{3}{8 \pi  G_N l^2}$. Although it is possible to define different pressures for the parameters $V_0$ and $\Lambda$, in our analysis, we assume that only $V_0$ exerts pressure on the black hole. In other words, we consider $V_0$ to be a thermodynamic variable. On the other hand, we treat $\Lambda$ as a constant that does not enter the first law of thermodynamics. This choice seems reasonable since the pressure associated with $V_0$ is a result of the dynamic process of spontaneous Lorentz symmetry breaking, which is well-established in the literature. This process is inherently dynamic, in contrast to the a priori nature of the cosmological constant.

%Although these two pressures will be important in obtaining some thermodynamic quantities, only the pressure $P_{LV}$ will be considered in the first law of thermodynamics, i.e., we will only consider that the LV massive mode exerts work on the black hole.
%In limit $z=1$ and $V_0 =0$ we recover the AdS pressure $P=\frac{3}{8 \pi  G_N l^2}$. From here,  we will no longer mention tangential pressures, since in the thermodynamic description only the radial pressure will be important, so that we will no longer use the sub-index $r$ to refer to it.
\section{Phase structure} \label{3}
In this section, we explore and analyze the critical behaviors of the black hole in Lifshitz spacetime (\ref{resumo1}). It is important to note that establishing an exact analysis of the $P-V$ extended phase space for Lifshitz solutions presents certain difficulties. This arises due to the challenge of finding an equation of state $P=P(V,T)$ for an arbitrary dynamic exponent $z$. Although this problem can be circumvented by assuming a modified equation of state, as seen in Ref.\cite{crit}, a more natural and convenient approach is to consider that the thermodynamic pressure is given by the LV (\ref{pr}), which simplifies the analysis of the criticality in the PV diagram. By adopting this choice, we can obtain an equation of state that closely resembles that of a Van der Waals fluid, as we will explore further.

Before delving into the analysis of the phase structure, it is essential to study the thermodynamic properties of the solutions (\ref{resumo1}-\ref{resumo5}) and verify if the first law is valid. However, to proceed, we must formally define how the parameter $m$ relates to the mass of the black hole. But first, let us explicitly define the LV charge from Eq. (\ref{carga2}) as:
\begin{equation} \label{car}
    \Tilde{Q} = \frac{\Tilde{q} l ^{z-1}}{4 G_N}.
\end{equation}
Further, the potential associated with this charge in the thermodynamic relations, measured at infinity with respect to the horizon, is defined by:
\begin{equation}
    \Tilde{\Phi}(r) = \Tilde{A}_{\mu}\chi^{\mu}|_{r\rightarrow\infty} -\Tilde{A}_{\mu}\chi^{\mu}|_{r=r_h}
\end{equation}
where $\chi=\partial_t$ is the null generator of the horizon.  Considering Eq.(\ref{resumo}) and the equation above, we obtain that
\begin{equation}\label{pot}
    \Tilde{\Phi}(r) =  - \frac{\Tilde{q}  e^{-\sqrt{z-1} \phi_0}}{z}(r^{-z}-r_h^{-z}),
\end{equation}

 Using the modified Brown and York formalism \cite{york}, we can calculate the mass of the solution as:
\begin{equation}\label{mass}
    M = \frac{m l^{-z-1}}{2 G_N }
\end{equation}
\subsection{Thermodynamics}
The first thermodynamic property that we calculate is the temperature, which can be given by $T=\frac{l^{-1-z} r^{z} \left(r f'+2 z f\right)}{4 \pi }$\footnote{We can calculate surface gravity $\kappa$, hence temperature $T= \frac{\kappa}{2\pi}$, for static spherically symmetric spacetime with metric given by
\begin{equation}
    ds^2 =  g_{tt}dt^2 + g_{rr}dr^2 + r^2d\Omega^2_{2},
\end{equation}
through the following formula
\begin{equation}
    \kappa = -\frac{\partial_r g_{tt}}{2\sqrt{-g_{rr} g_{tt}}}
\end{equation}}. By substituting Eqs. (\ref{resumo2}) and (\ref{massa}) into this expression, we find that:
\begin{equation}\label{temp}
    T = \frac{l^{-z-1} r_h^{-z-2} \left[4 r_h^{2 z} \left(l^2+r_h^2 z (z+2)\right)-z e^{-\sqrt{z-1} \phi_0} \left(\Tilde{q}^2 l^{2 z}+2 l^2r_h^4 V_0\right)\right]}{16 \pi  z},
\end{equation}
where we use the Eq. (\ref{mass}). Thus, the extremal limit ($T=0$) is given by
\begin{equation}\label{ext}
  V_0^{ext} = \frac{4 e^{\sqrt{z-1} \phi_0} r_{ext}^{2 z} \left(z (z+2) r_{ext}^2+l^2\right)-\Tilde{q}^2 z l^{2 z}}{2 l^2 z r_{ext}^4},
\end{equation}
 where $r_{ext}$ is the position of the horizon at extremality. Notably, if we substitute Eqs. (\ref{massa}) and (\ref{ext}) into (\ref{resumo2}), we indeed find that the near-horizon geometry is described by $AdS_{2} \times S^2$. % More, it is possible to calculate the extremal mass value as well, just substitute Eq. (\ref{ext}) in Eq(\ref{massa}), so that we have
%  \begin{equation}
 %     m_{ext} = \frac{r_{ext}^{-z} \left[\frac{2 l^2 (z-2) r_{ext}^{2 z}}{z^2}+2 (z-1) r_{ext}^{2 z+2}-\frac{\Tilde{q}^2 l^{2 z} e^{-\sqrt{z-1} \phi_0}}{z}\right]}{z-4}
 % \end{equation}

The other quantity is the Bekenstein-Hawking entropy which is given by
\begin{equation}\label{entropia}
    S = \frac{\pi r_h^2}{G_N}.
\end{equation}

Having defined the thermodynamic and geometric quantities for the solution (\ref{resumo1}), we are now ready to verify the first law of thermodynamics. However, before proceeding, it is essential to understand the role of the parameters of the Lorentz violation, expressed by $V_0$ and $\Tilde{q}$, in the first law. Inspired by the limit $z \rightarrow 1$, we can assume that the cosmological constant is fixed, meaning the black hole parameters are varied in a 'fixed AdS background'. On the other hand, as already mentioned, we consider that the LV massive mode contributes to a form of pressure in the first law. Drawing parallels between $V_0$ and the cosmological constant, we will adopt the analysis made by Refs \cite{entalpia1,entalpia2} to understand the implications of $V_0$ in the first law of thermodynamics. Indeed, these articles demonstrated that when the cosmological constant is considered as a thermodynamic variable through a pressure term, the mass of the black hole is no longer identified solely with the free energy, but rather with the \textit{enthalpy}. This novel approach garnered considerable attention as it unveiled the striking similarities between the AdS charged black hole solutions and the Van der Waals fluid \cite{ads}, providing a clearer and more insightful perspective on their thermodynamic behavior.  Certainly, this assumption does raise some concerns, notably the lack of a known mechanism that treats the cosmological constant as vacuum expectation values. However, for our proposal, this does not pose a problem, as the bumblebee excitations generating these new terms arise precisely as vacuum expectation values of $B_{\mu}$. This unique feature sets our approach apart, allowing for a natural incorporation of the cosmological constant and avoiding the issues associated with the traditional treatment of it as a thermodynamic variable.

Therefore, by assuming that the LV mass mode $V_0$ is a thermodynamic variable, we establish that the mass parameter (\ref{mass}) is analogous to the enthalpy. This thermodynamic potential can then be related to the entropy, pressure, and charge in an extended phase space that includes the $P_{LV}$ and $V_{LV}$ variables. The relationship can be expressed as follows:
\begin{equation}\label{law}
    dH = TdS + V_{LV} dP_{LV} + \Tilde{\Phi}(\infty)d\Tilde{Q},
\end{equation}
where the potential $\Tilde{\Phi}$ is held fixed at the boundary at value  $\Tilde{\Phi}(\infty) = \frac{\Tilde{Q} l^{z-1} r_h^{-z} e^{-\sqrt{z-1} \phi_0}}{4 G_N z} $, and serves as the variable conjugate to the charge $\Tilde{Q}$. Thus, we have that
\begin{equation} \label{rel1}
    T = \bigg( \frac{\partial H}{\partial S} \bigg)_{V_0, \Tilde{Q}, \phi _0}
\end{equation}
\begin{equation}\label{rel2}
    V_{LV} = \bigg(\frac{\partial H}{\partial P}\bigg)_{S,\Tilde{Q},\phi_0}.
\end{equation}\label{rel3}
\begin{equation}\label{rel3}
   \Tilde{\Phi}(\infty)  = \bigg(\frac{\partial H}{\partial \Tilde{Q}}\bigg)_{S,V_0,\phi_0}.
\end{equation}
Indeed, it is evident that the relation (\ref{rel1}) holds, as the enthalpy is given by Eq. (\ref{mass}). Similarly, we find that Eq. (\ref{rel3}) holds for (\ref{car}). Finally, we can confirm that the first law truly holds when the "thermodynamic volume" (\ref{rel2}) is given by:
\begin{equation}\label{volume}
V_{LV} = \frac{4 \pi l^{1-z} r_h^{z+2}}{4-z},
\end{equation}
where we assume that the thermodynamic pressure is given by the pressure (\ref{pr}). Indeed, quantity (\ref{volume}) has volume dimension, $[length]^3$.  In limiting cases, our expressions naturally recover the definitions in \cite{kubizvnak2012p,ads1,ads2}.

%$V$ is  the thermodynamic volume given by $(\frac{\partial H}{\partial P})_{S,\phi_0}$. %Similarly, we have that the temperature is given by $(\frac{\partial H}{\partial S})_{P,\phi_0}$. It is straightforward to verify that assuming the volume given by
%\begin{equation}\label{volume}
%V = \frac{4}{3}\pi r_h^3,
%\end{equation}
%the first law (\ref{law}) holds, if the pressure exerted by the LV in the formation of the black hole asymptotically Lifshitz is given by
%\begin{equation}\label{pressa}
 %   P = - \frac{3 e^{-\sqrt{z-1}\phi_0}l^{1-z}r_h^{1-z}}{16\pi G_N(4-z)}V_0.
%\end{equation}
%Note that in the limit $z=1$, we have a negative pressure given by $P = -\frac{V_0}{16\pi G_N}$, i.e., very similar to that obtained by Ref. \cite{pv}. {\color{red}{A RELAÇÃO DE SMARR É VÁLIDA?}} 

Substituting Eq.(\ref{entropia}) and (\ref{pr}) in the expression (\ref{massa}), we obtain the enthalpy given by
\begin{equation}
    H(S, P_{LV},\Tilde{Q}) =  \frac{\pi ^{z/2}  \left(-\frac{16 G_N^3 l^2 P_{LV} S^2 z^2 r_h^{2 z-2}}{\pi  (z-4)}+8 G_N^2 l^2 \Tilde{Q}^2 z e^{-\sqrt{z-1} \phi_0}+2 \pi ^{-z} \left( S G_N\right)^{z} \left(\frac{G_N S z^2}{\pi }+l^2\right)\right)}{4 G_N z^2 l^{z+1} \left( \sqrt{S G_N}\right)^{z}}
\end{equation}
In fact,  the Eq. (\ref{law}) holds, since the quantities (\ref{rel1}), (\ref{rel2}) and (\ref{rel3}) are confirmed with Eqs. (\ref{temp}), (\ref{volume}) and (\ref{pot}), respectively. 
In addition, in the limit $z=1$ ($\Tilde{Q}=0$) we recover the enthalpy found by Ref.\cite{entalpia1}, as expected.

\subsection{Equation of state}
In order to explore the similarities between the Lorentz Violation black hole solution in Lifshitz spacetime (\ref{resumo1}) and the Van der Waals fluid, we make the assumption that both the charge and the cosmological constant are fixed external parameters, not thermodynamic variables. To proceed in this direction, we obtain the black-hole equation of state, i.e., the equation that depends only on $P_{LV}$ and $V_{LV}$, from Eq. (\ref{temp}), and it is given by:
\begin{equation}\label{eoe}
   P_{LV}(T,V_{LV}) =  -P_{\Lambda_0}+\frac{T l^{z-1} r_h^{-z}}{2 G_N}+\frac{G_N \Tilde{Q}^2 r_h^{-2 (z+1)} e^{-\sqrt{z-1} \phi_0}}{2 \pi }-\frac{1}{8 \pi  G_N r_h^2 z},
\end{equation}
%\begin{equation}
 %   T(P,V)=\frac{l^{1-z} \left[ 2^{3-\frac{4}{z+2}} \pi ^{\frac{z}{z+2}} G_N P  \left[V (z-4) \left(-l^{z-1}\right)\right]^{\frac{2}{z+2}}+1\right]}{4 \pi \left[(4 \pi )^{-\frac{1}{z+2}} \left[V (z-4) \left(-l^{z-1}\right)\right]^{\frac{1}{z+2}}\right]^{2-z}}.
%\end{equation}
where the $V_{LV}$ is given in terms of the event horizon radius $r_h$ through Eq.(\ref{volume}). Also, $P_{\Lambda_0}$ is defined in (\ref{presslam}), but to reiterate, this quantity is a fixed parameter. Once we have the equation of state (\ref{eoe}), we can proceed to calculate the critical points of the P-V diagram. However, before that, it would be more appropriate to rewrite Eq. (\ref{eoe}) in a form similar to the equation of state for the Van der Waals fluid \cite{ads}. To achieve this, we define a kind of "specific volume" given by:
\begin{equation}
    \mathrm{v} \equiv \bigg(\frac{2G_N}{l^{z-1}}\bigg)r_h^z.
\end{equation}
Thus, we have Eq. (\ref{eoe}) can be rewritten as
\begin{equation} \label{estado}
    P_{LV}  = -\frac{a}{2 \pi   \mathrm{v}^{2/z}}+\frac{b}{\pi   \mathrm{v}^{\frac{2 (z+1)}{z}}}+\frac{T}{ \mathrm{v}} - P_{\Lambda_0},
\end{equation}
where
\begin{equation}
    a=\frac{2^{\frac{2}{z}-4} G_N^{\frac{2}{z}-1} l^{\frac{2}{z}-2}}{z}
\end{equation}
\begin{equation}
    b=2^{\frac{z+2}{z}}\Tilde{Q}^2  G_N^{\frac{2}{z}+3} l^{\frac{2}{z}-2 z} e^{-\sqrt{z-1} \phi_0}.
\end{equation} Moreover, we can define the "specific volume" as $\mathrm{v}=\frac{V_{LV}}{N}$, where $N$ represents the "number of states" \cite{Kubiznak:2016qmn}. Thus, it is straightforward to obtain that for (\ref{volume}), the general $N$ is given by:
\begin{equation}
    N = \frac{2}{4-z} S,
\end{equation}where $S$ is the entropy (\ref{entropia}). Indeed, this remarkable result indicates that the entropy of the black hole is somehow related to the number of degrees of freedom of the system. The fact that the expression for $N$ is proportional to the black hole entropy suggests a deeper connection between the microscopic properties of the black hole and its macroscopic thermodynamic behavior, providing valuable insights into the underlying nature of the black hole solution.

\begin{figure}[!ht] 
      \includegraphics[height=4.8cm]{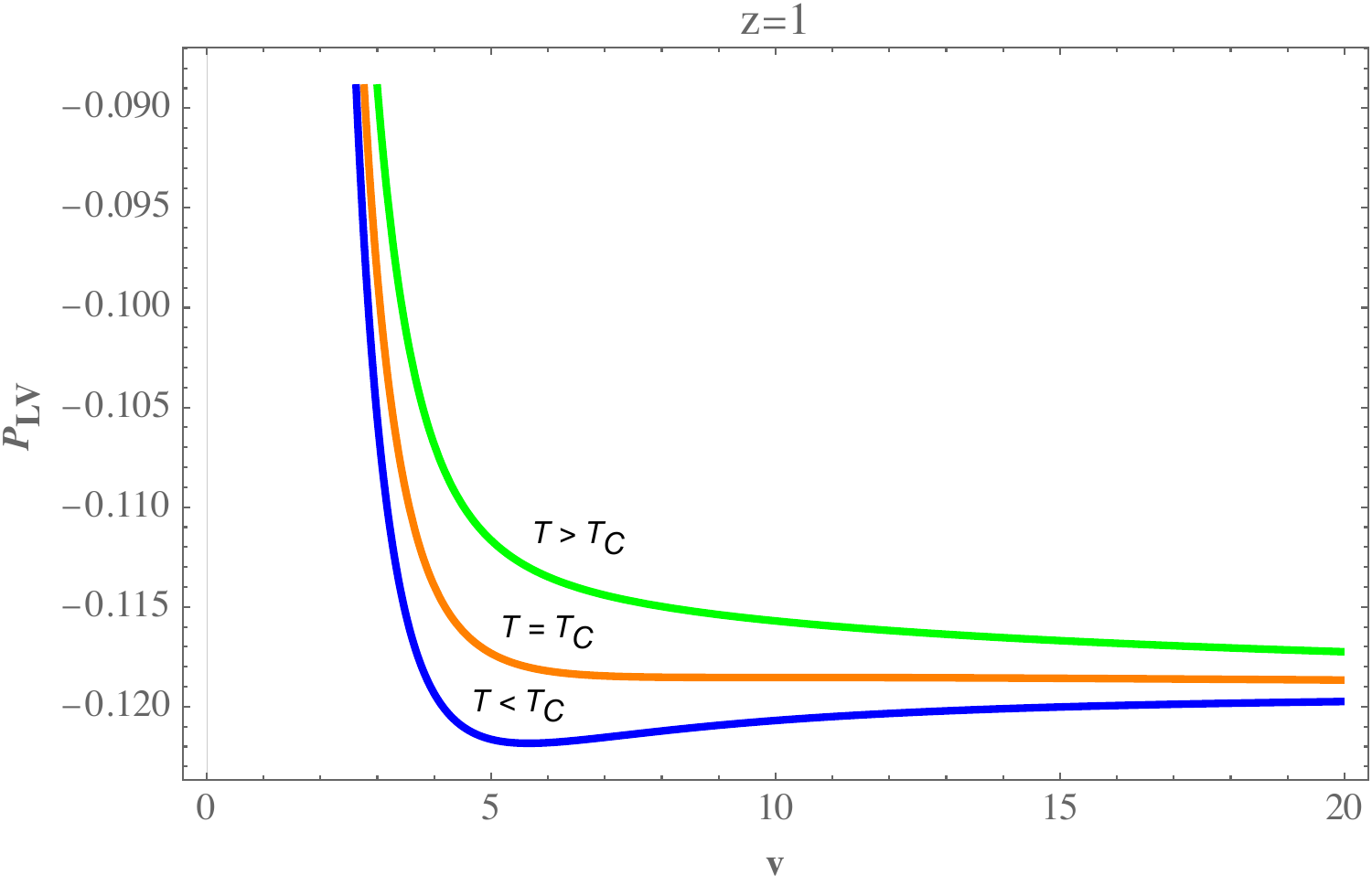}\quad
      \includegraphics[height=4.8cm]{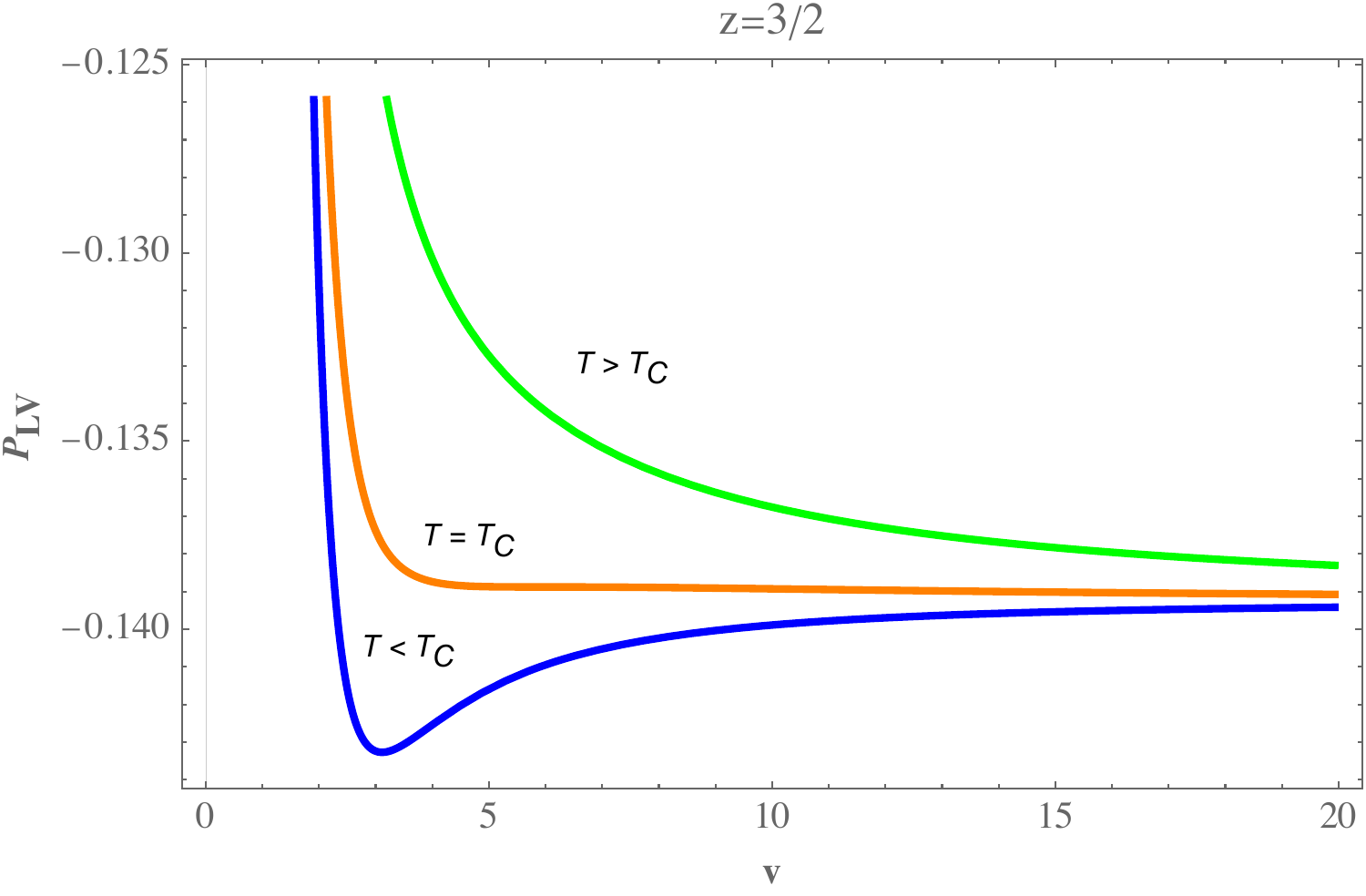}\quad
    \caption{P-V diagram of Einstein-Maxwell-Bumblebee-Dilaton black hole for $\Tilde{Q}=1$. We have set $G_N=\phi_0=l^2=1$}  \label{dia}
\end{figure}

 %let us first define an effective pressure that takes into account both the pressure due to LV and the cosmological constant, i.e.
%\begin{equation}
 %   P_{eff} \equiv P_{LV} + P_{\Lambda_0}.
%\end{equation}
The critical point is obtained from
\begin{equation}
   \bigg( \frac{\partial P_{LV}}{\partial   \mathrm{v}} \bigg)_{T, \Tilde{Q}, \phi _0}=\bigg(  \frac{\partial^2 P_{LV} }{\partial   \mathrm{v}^2}\bigg)_{T, \Tilde{Q}, \phi _0} = 0,
\end{equation}
which leads to
\begin{equation}\label{volc}
     \mathrm{v}_c = 4 G_N^2 \Tilde{Q} l^{1-z} e^{-\frac{\sqrt{z-1} \phi_0}{2}} \sqrt{\frac{z (z+1) (z+2)}{2-z}} 
\end{equation}
\begin{equation}\label{temc}
    T_c = \frac{2^{\frac{z-2}{z}}G_N^{\frac{z-2}{z}} l^{1-z} \Tilde{Q}^{\frac{z-2}{z}} e^{-\frac{(z-2) \sqrt{z-1} \phi_0}{2 z}}\left(\frac{z (z+1) (z+2)}{2-z}\right)^{\frac{1}{2}-\frac{1}{z}}}{\pi  z (z+2)}
\end{equation}
\begin{equation}\label{prec}
    P_c = \frac{2^{-\frac{2}{z}-3} G_N^{-\frac{z+2}{z}} \Tilde{Q}^{-2/z} e^{\frac{\sqrt{z-1} \phi_0}{z}}(z+2) \left(\frac{z (z+1) (z+2)}{2-z}\right)^{-\frac{z+1}{z}}}{\pi } - P_{\Lambda_0}.
\end{equation}
From the equations presented above, it becomes evident that to identify critical points, where both the critical temperature and the critical horizon are real and non-negative, the dynamic coefficient needs to be within the range of $1\leq z<2$. By plotting two P-V diagrams for two values of $z$ within this interval, we observe the emergence of an 'oscillating part' in the isotherm, closely resembling the Van der Waals P-V diagram (Fig. \ref{dia}). Notably, the pressure near the critical point is negative for this configuration. This is a characteristic feature of scenarios where the Lorentz symmetry is spontaneously broken (\ref{pr}) for $V_0>0$. However, as the temperature increases, we move away from the critical points, until we reach a positive pressure $P_{LV}$, indicating a scenario where the spontaneous breaking of the Lorentz symmetry is no longer present, as we would have $\lambda<0$. %However, we could find a temperature, or in the more appropriate case a charge (note in Eq. (\ref{temc}) that the temperature is inversely proportional to the bumblebee charge), whose Lorentz symmetry is restored. For this purpose, it suffices that $P_c\geq 0$, i.e.,
%\begin{equation}
 %   \Tilde{Q}\geq\sqrt{\alpha}P_{\Lambda_0}^{-z/2}
%\end{equation}
%where
%\begin{equation}
 %   \sqrt{\alpha}\equiv \frac{(2-z)e^{\sqrt{z-1} \phi_0}}{G_N^{2}z(z+1)(z+2)}\bigg[\frac{(2-z)}{2^{\frac{3z+2}{3}} \pi G_N z(z+1)} \bigg]^z
%\end{equation}
%However, it is straightforward to calculate that the above equations do not admit critical points. Furthermore, we could not observe any van der Waals like behavior, as we can confirm in Fig. (\ref{dia}). Given all this, our solution is expected to be thermodynamically stable. In fact, we can confirm this by calculating  the specific heat

We can also find a universal relation for general $z$ that relates the three critical quantities (\ref{volc}),(\ref{temc}) and (\ref{prec}). Thus, it is straightforward to show that
\begin{equation}
    \frac{P_c^{eff} \mathrm{v}_c}{T_c}= \frac{4-z^2}{4(1+z)}
\end{equation}
where $P_c^{eff} = P_c + P_{\Lambda_0} $. Note that in $z=1$, we recover same relation as for the Van der Waals fluid. 
%In order to obtain the predicted universal relation for the Van der Waals fluid for Lifshitz solution, i.e., $\frac{P_c^{eff} \mathrm{v}_c}{T_c}=3/8$, where $P_c^{eff} = P_c + P_{\Lambda_0} $, the dilaton amplitude $\phi_0$ needs to be given by
%\begin{equation}\label{ampl}
%    \phi_0 = \frac{ln[G^4_N\Tilde{Q}^2 l^{2(1-z)}\chi(z)]}{\sqrt{z-1}},
%\end{equation}
%where
%\begin{equation}
   % \chi(z) \equiv \frac{3^{\frac{2z}{z-1}}z(1+z)(2+z)}{2-z} %\bigg[ \frac{2^{\frac{2-z}{z}}(2+z)(2-z)}{1+z} \bigg].
%\end{equation}
%This choice for $\phi_0$ cancels precisely the parameters of the theory in the universal relation. In addition, we have Eq. (\ref{ampl}) is not well defined in $z=1$.
Further, from the equation of state (\ref{estado}) it is possible to obtain 'the law of corresponding states'. Setting the following quantities
\begin{equation}
    p = \frac{P^{eff}}{P_c^{eff}} , \  v = \frac{\mathrm{v}}{\mathrm{v}_c} , \ \tau = \frac{T}{T_c},
\end{equation}
where $P^{eff} = P_{LV}+P_{\Lambda_0}$, we have that
\begin{equation} \label{law}
    8\tau = 2(2+z)v\left[ \frac{2-z}{1+z}p +\frac{1}{v^{\frac{2}{z}-1}} \right] - \frac{2(2-z)}{(1+z)v^{-1-\frac{z}{2}}}.
\end{equation}
In limit $z=1$, we recover 'the law of corresponding states' of Ref. \cite{ads}. This equation will be crucial for calculating the critical exponents later on.

\subsection{Thermal stability}\label{c}
\begin{figure}[!ht] 
 \centering
      \includegraphics[height=5cm]{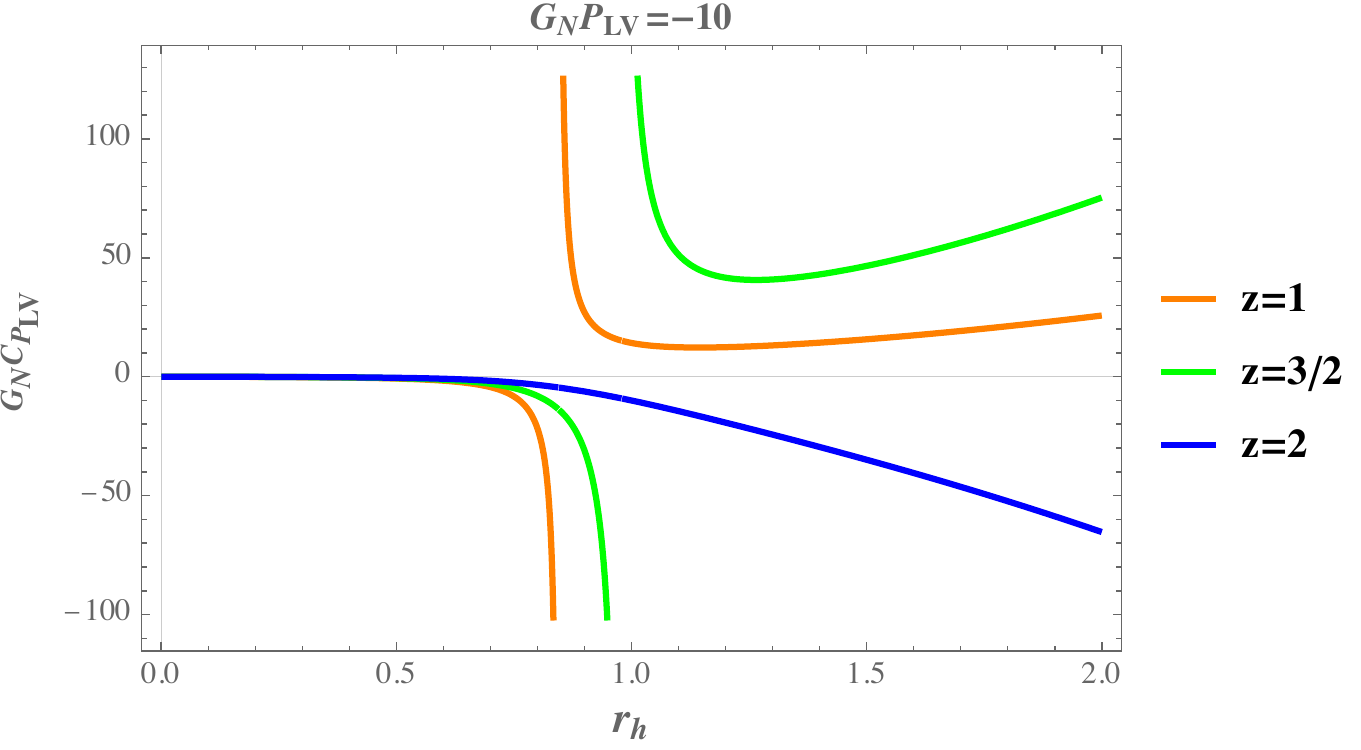}\quad
    \caption{Heat capacity at constant pressure vs. horizon radius for various values of $z$ and $G_N \Tilde{Q}=1$. We have set $\phi_0=l^2=1$}  \label{diacp}
\end{figure}

As shown by Ref.\cite{chamblin1999holography} , it is possible to associate thermodynamic stability with microscopic fluctuations of the system.  The stability condition can be expressed as
\begin{equation} \label{estab}
    C_{P,V} \equiv T \bigg(\frac{\partial S}{\partial T}\bigg)_{P_{LV},V_{LV},\Tilde{Q}} \geq 0  
\end{equation}
where $C_{P,V}$ is the specific heat at constant pressure or volume.  Using Eqs. (\ref{temp}) and (\ref{entropia}), we have that the specific heat at constant pressure is given by
\begin{align} \label{cp} \nonumber
     &C_{P} =  \frac{ 8 \pi ^2  z l^{z+1} r_h^{z+4} e^{\sqrt{z-1} \phi_0}T}{ \Tilde{B}}
\end{align}
where $T$ is given by (\ref{temp}) and 
%\begin{equation}
 %   \Tilde{A}\equiv r_h^{2 z} e^{\sqrt{z-1} \phi_0} \left[l^2 \left(8 \pi  G_N P r_h^2 z+1\right)+r_h^2 z (z+2)\right]-4 G_N^2 l^2 \Tilde{Q}^2 z
%\end{equation}
\begin{equation}
    \Tilde{B}\equiv 4 G_N^3 l^2 \Tilde{Q}^2 z (z+2)+G_N r_h^{2 z} e^{\sqrt{z-1} \phi_0} \left[r_h^2 z^2 (z+2)-l^2 (z-2) \left(8 \pi  G_N P_{LV} r_h^2 z-1\right)\right]
\end{equation}
Immediately, we can observe that the specific heat $C_{P}$ becomes singular at $\Tilde{B}=0$, precisely at the critical point. In Fig. (\ref{diacp}), we plot the heat capacity at constant pressure against the horizon, considering various values of $z$. The orange and green lines exhibit discontinuities, which is a result of these lines falling within the range allowing for critical points $1\leq z<2$. The blue line, situated outside this interval, does not display discontinuities. Furthermore, we can observe that only the two lines within the interval demonstrate thermodynamic stability. On the other hand, the heat capacity at constant volume vanishes, as the entropy (\ref{entropia}) remains unchanged when the volume is fixed.

%the specific heat at constant pressure does not show divergences, as expected. Furthermore, we can conclude that stability is achieved for $r_h>r_0$, if we assume that $P\geq0$, where $r_0$ is called bound point. This quantity is obtained for $\Tilde{B}=0$, so that it can be expressed by
%\begin{equation}\label{point}
 %   r_0 = \frac{l | z-2| }{\sqrt{2z \left[-4 \pi  G l^2 P% (z-2)+z^2+z-2\right]}}.
%\end{equation}
%However, it is still necessary that the denominator of Eq. (\ref{point}) be positive. Thus, we have that the condition (\ref{estab}) is valid if
%\begin{equation}\label{inter}
 %   0 \leq P \leq \frac{z^2+z-2}{4 \pi G_N l^2 (z-2)}
%\end{equation}

%\subsection{Grand-canonical ensemble}
%\begin{equation}
 %   G(T,P) = \frac{r_h^2 \left\{l^{-z-1} r_h^z \left[\frac{z \left[r_h^2 z (z+2)-l^2 \left(8 \pi  G_N P r_h^2 z+z-1\right)\right]+l^2 (z-4)}{r_h^2 z^2 (z-4)}+1\right]+2 \pi  T\right\}}{2 G_N}
%\end{equation}
%\subsection{Canonical ensemble}
%\begin{equation}
 %   F(T,V) = \frac{l^{-z-1} \left\{-2 \pi  r_h^2 T (z-4) z^2 l^{z+1}-l^2 [(z-2) z+4] r_h^z+2 (z-1) z^2 r_h^{z+2}\right\}}{2 -N (z-4) z^2}
%\end{equation}
\subsection{Gibbs free energy}

In phase transition studies, analyzing the Gibbs free energy is of utmost importance. This is because first-order transitions are identified by the divergent behavior of its first derivative. With this in mind, we can express the Gibbs free energy as follows:
\begin{equation}
    G = H - T S,
\end{equation}
where the enthalpy $H$ is identified as the mass (\ref{mass}), as discussed earlier. Thus, we have that the Gibbs energy for fixed $\Tilde{Q}$ for our solution is given by
\begin{align}
    G(T,P_{LV})= \frac{ \left[l^2 \left(4 G_N^2 \Tilde{Q}^2 z (z+2) e^{-\sqrt{z-1} \phi_0}-\frac{(z-2) r_h^{2 z} \left(8 \pi  G_N P_{LV} r_h^2 z^2+z-4\right)}{z-4}\right)-z^3 r_h^{2 z+2}\right]}{4 G_N z^2 l^{z+1} r_h^{z}}
\end{align}
\begin{figure}[!ht] 
      \includegraphics[height=4.8cm]{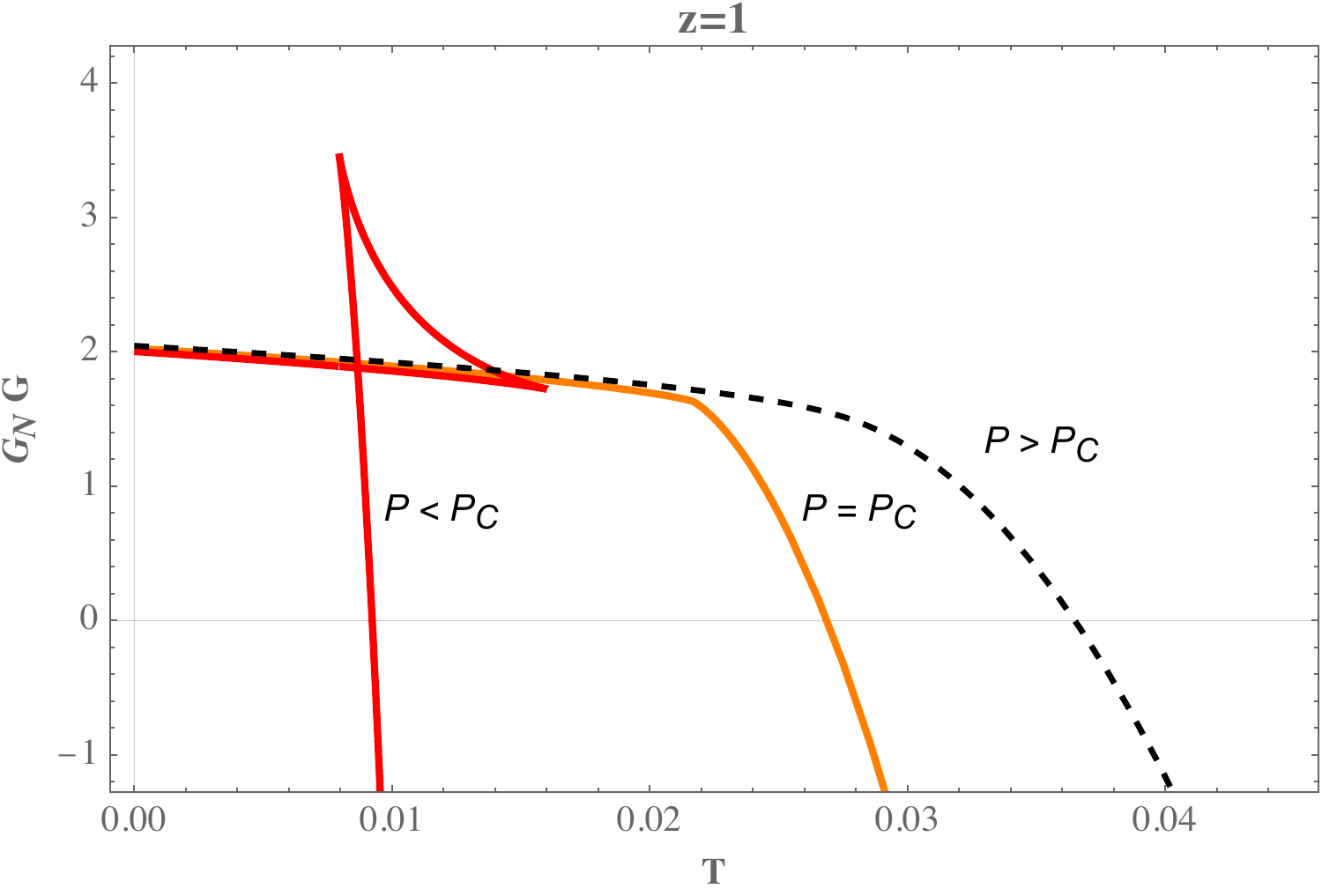}\quad
      \includegraphics[height=4.8cm]{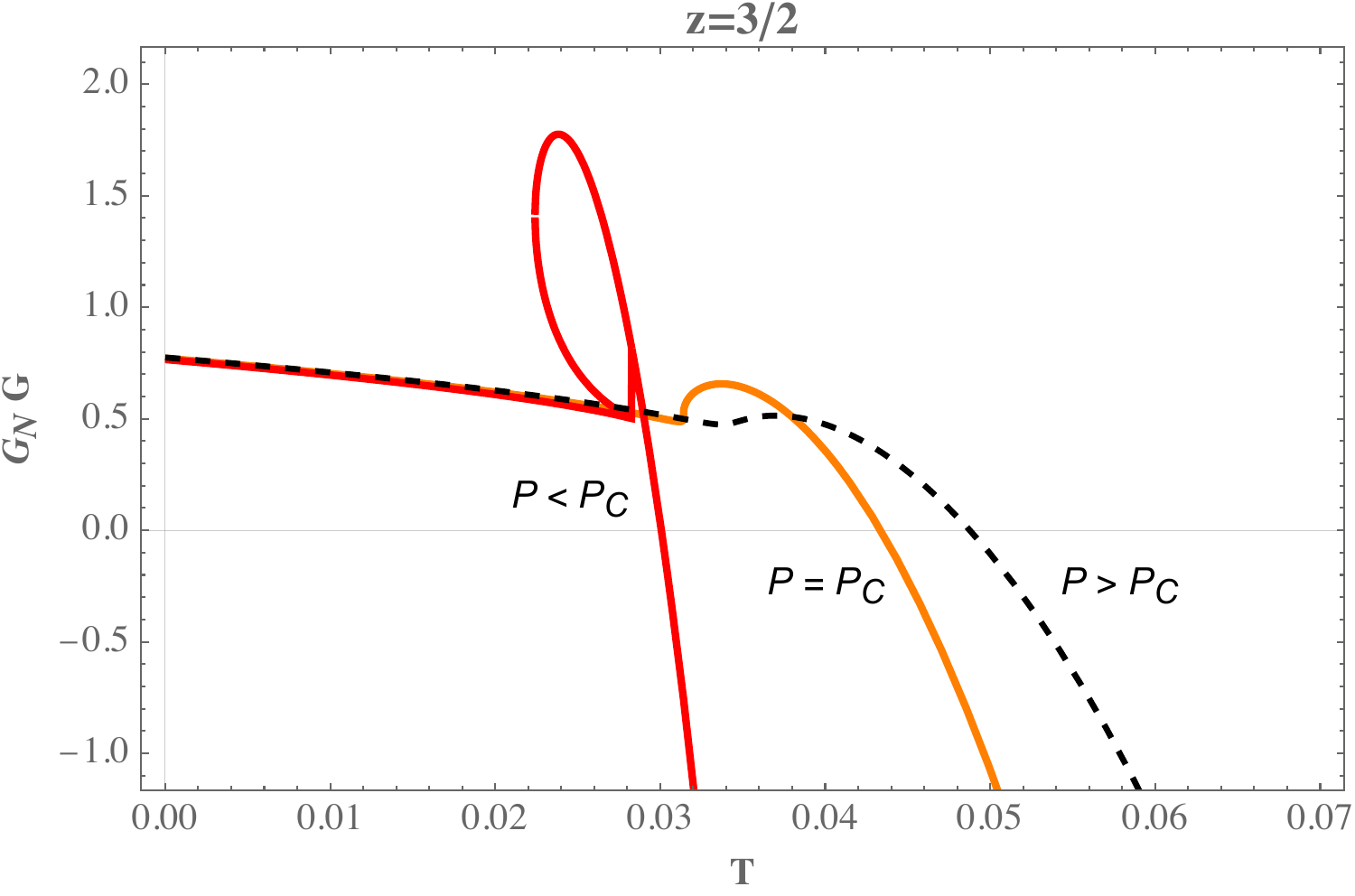}\quad
    \caption{Gibbs free energy of Einstein-Maxwell-Bumblebee-Dilaton black hole for $G_{N}\Tilde{Q}=1$. The orange lines correspond to critical pressures where $G_{N}P_c^{z=1}\approx -0.118 $  and $G_{N}P_c^{z=3/2}\approx -0.139 $ .We have set $\phi_0=l^2=1$}  \label{GIBBS}
\end{figure}
where $r_h=r_h(T,P_{LV}) $ is expressed through Eq.(\ref{eoe}). In the limit $z=1$ we retrieve the same result from Ref.(\cite{kubizvnak2012p}) that used the Euclidean approach to quantum gravity to obtain the same result.

In Fig. (\ref{GIBBS}), we present the Gibbs free energy as a function of temperature for two values of $z$ that fall within the range allowing for critical points. For $z=1$, we observe a graph very similar to the one obtained in \cite{kubizvnak2012p}. In fact, we can see that for $P<P_{c}$, the "swallow tail" phenomenon emerges. On the other hand, when we set $z=3/2$, the "swallow tail" behavior is altered. However, for both cases, it is evident that when $T<T_c$, a first-order phase transition occurs between a (small black hole) and a (large black hole).

\subsection{Critical exponents }
The behavior of physical quantities near a critical point can be characterized by critical exponents. These exponents are independent of the details of the theory, making them universal. However, they may depend on the size of the system or the range of interactions. Using the same approach as in Ref. \cite{ads}, we obtain four critical exponents: $\alpha$, $\beta$, $\gamma$, and $\delta$ below.
\begin{itemize}
    \item \emph{Order parameter $\eta$}
\end{itemize}
In order to obtain the order parameter, we expand the equation of corresponding states (\ref{law}) in the neighborhood of the critical point. For this, we define that:
\begin{equation}
    t = \tau -1  \ \  \omega = v -1.
\end{equation}
So, we obtain that:
\begin{equation}\label{aprox}
p \approx 1-\frac{4(z+1)t}{z^2-4} +\frac{4z(z+1)t\omega}{z^2-4}-\frac{2z(z+1)\omega^3}{3} + \dots,
\end{equation}
where the ellipsis takes into account the terms of order equal to or greater than $\omega^4$, which we disregard. We also neglect the $t\omega^2$ terms. Now, we can calculate the differential of the truncated series above, and we get that:
\begin{equation}
    d P_{LV} = 2  z (z+1) P_c^{eff} \left(\frac{2 t}{z^2-4}-\omega ^2\right) \text{d$\omega $}.
\end{equation} It is worth remembering that $P_{\Lambda_0}$ is fixed, so its differential vanishes. Assuming that $\omega_l$ and $\omega_s$ are the 'volumes' of the large and small black holes, respectively, and that during the phase transition the pressures are equal, we can obtain the following equation:
\begin{equation}\label{equ1}
    1-\frac{4  (z+1)t}{z^2-4} +\frac{4 z (z+1) t \omega_l }{z^2-4}-\frac{2 z (z+1)\omega_l ^3}{3} = 1-\frac{4  (z+1)t}{z^2-4} +\frac{4 z (z+1) t \omega_s }{z^2-4}-\frac{2 z (z+1)\omega_s ^3}{3}.
\end{equation} Furthermore, we can employ the Maxwell’s equal area law to obtain the following equation:
\begin{equation}\label{equ2}
    \int_{\omega_l}^{\omega_s} \omega \left(\frac{2 t}{z^2-4}-\omega ^2\right) \text{d$\omega $} = 0.
\end{equation} Assuming fixed $t<0$, the unique non-trivial solution to the system of equations (\ref{equ1}) and (\ref{equ2}) is $\omega_l=-\omega_s\propto \sqrt{-t}$. Thus, the order parameter is given by
\begin{equation}
    \eta = \mathrm{v}_{c}(\omega_l-\omega_s)\propto\sqrt{-t}.
\end{equation} 
Therefore, we find that the first critical exponent is $\beta = 1/2$.

\begin{itemize}
    \item \emph{Isothermal compressibility $\kappa_T$}
\end{itemize}
The other critical exponent $\gamma$ is found through the isothermal compressibility given by
\begin{equation}
    \kappa_T = - \frac{1}{v}\frac{\partial v}{\partial P_{LV}}\bigg|_{T} \propto t^{-\gamma}.
\end{equation}
We achieve this by differentiating Eq. (\ref{aprox}). Thus, we can easily obtain that
\begin{equation}
    \kappa_T  \propto \frac{1}{P_c^{eff}t}.
\end{equation}
Therefore, we find that the second critical exponent is $\gamma = 1$.

\begin{itemize}
    \item \emph{Critical isotherm $T=T_c$}
\end{itemize}
The next critical exponent $\delta$ is related to the ‘shape of the critical isotherm’ $t=0$ through the relation:
\begin{equation}
|P-P_c|\propto|v-v_c|^\delta.
\end{equation}
We can obtain this relation by assuming $t=0$ in Eq.(\ref{aprox}). Thus, we get:
\begin{equation}
p-1 = -\frac{2 z (z+1)\omega ^3}{3}.
\end{equation}
Therefore, we find that the third critical exponent is $\delta = 3$.

\begin{itemize}
    \item \emph{ Specific heat at constant volume $C_{\mathrm{v}}$}
\end{itemize}
Finally, we have that the exponent $\alpha$ governs the behaviour of the specific heat at constant volume,
\begin{equation}
    C_{\mathrm{v}} = T \frac{\partial S}{\partial T}\bigg|_{\mathrm{v}}\propto |t|^{\alpha}.
\end{equation}
But as mentioned at the end of subsection \ref{c}, the $C_{\mathrm{v}}$ vanishes. Therefore, we find that the last critical exponent is $\alpha = 0$.

In conclusion, we have found that the critical exponents for a dilaton Lifshitz black hole with bumblebee excitations are the same as those found in the Van der Waals fluid. This remarkable similarity indicates that the critical behaviors in both systems share universal features, despite their different physical origins. The study of critical phenomena in this black hole solution provides valuable insights into the thermodynamic properties of systems with Lifshitz symmetry and Lorentz violation.

\section{Final remarks and perspectives } 
\label{con}
We investigated the impact of the dilaton coupling on multiple gauge fields and bumblebee excitations in (3+1)-dimensional asymptotically Lifshitz black hole solutions. In the context of a static and spherically symmetric black hole, the transverse and longitudinal modes become decoupled from the radial vacuum expectation value (VEV). Subsequently, we developed a model that links the dilaton to these modes in a similar fashion. Moreover, the dilaton potential is induced by the massive mode $\beta_0$, while the massless mode of Nambu-Goldstone gives rise to a Maxwell-type field.  Ultimately, we demonstrated that the gauge fields function as auxiliary fields in our model. Furthermore, we ensured the decoupling of the cosmological constant term from the dilaton to guarantee the asymptotic Lifshitz behavior of our solution.

We have discovered a novel charged black hole solution within an asymptotically Lifshitz spacetime. When $z=1$, this solution reverts to the AdS charged black hole solution, albeit generated in the framework of spontaneous Lorentz symmetry breaking. For $z\neq 1$, we derived a solution displaying a complex phase structure that is highly contingent on the critical exponent.

To uncover the parallels between our LV solution and the Van der Waals system, we employed a criticality analysis in the $P-V$ space. Here, we treated the massive mode as the thermodynamic pressure and its conjugate quantity as the thermodynamic volume. Notably, we recognized that this pressure assumes a negative value, a necessary condition for the spontaneous breaking of the Lorentz symmetry, leading to the emergence of the $\beta_0$ mode and its association with a de Sitter phase. While it is conceivable to elevate the cosmological constant to a thermodynamic pressure, effectively representing the Anti de Sitter phase, we specifically elevated $V_0$ to a thermodynamic variable. This deliberate choice allowed for a focused examination of the solution's critical behavior through the lens of the $P-V$ diagram. We justified this decision by considering the ongoing dynamics of the massive longitudinal mode of the LV, which arises from fluctuations around the vacuum expectation value $b_{\mu}$. In this way, we provided a dynamic rationale for our selection

In the concluding section of the paper, our focus shifted towards a comprehensive analysis of the phase structures. We initiated this examination by deriving key thermodynamic quantities. First and foremost was the temperature. We established that in the extremal limit ($T=0$), the near-horizon geometry can be accurately described as $AdS_{2}\times S^2$. Additionally, we computed the entropy employing the Bekenstein-Hawking formula. Moreover, we underscored the necessity, within the framework of an extended $P-V$ phase space, to interpret the mass of the black hole as the enthalpy for the first law of thermodynamics to hold. However, the most pivotal aspect of this endeavor lies in the critical behaviors we unearthed during our scrutiny of the phase structures

We determined the critical points of the equation of state and constructed the $P-V$ phase diagram. In the range $1\leq z<2$, we observed the emergence of an 'oscillating component' in the isotherm, akin to our findings in the Van der Waals system. Additionally, we uncovered a universal relation and 'the law of corresponding states' dependent on $z$ that links the critical points. Notably, at $z=1$, we retrieve the relationships for the Van der Waals fluid. Concerning the stability of the solution, we demonstrated that $C_{P}$ exhibits discontinuities (indicative of phase transitions) and positivity (indicative of thermodynamic stability) for $1\leq z<2$. Moreover, the swallowtail behavior of the Gibbs free energy at $z=1$ signifies a first-order phase transition within the system. Conversely, for $z=3/2$, we observed a distinct swallowtail behavior, which also signals a first-order transition. Finally, we calculated the critical exponents of the system and determined them to be universal, mirroring those of a Van der Waals fluid system

To conclude the paper, we will offer some suggestions regarding the future directions and perspectives for this research. 
An intriguing avenue to explore would involve interpreting this in the context of dual field theory. Specifically, leveraging the AdS/CFT correspondence could provide insights into the non-gravitational manifestation of this system. 
In cases where z takes on arbitrary values, the system exhibits Lifshitz scaling. The corresponding dual boundary field theory, while not adhering to relativistic principles, still permits particle production. Therefore, it would be promising to understand how LV relates to a non-relativistic dual field theory. An additional proposition would involve delving into the potential identification of the Bumblebee field as the infrared limit of an extended version of Hor\v{a}va gravity \cite{PhysRevLett.104.181302}.
\acknowledgments
The authors thank the Conselho Nacional de Desenvolvimento Científico e Tecnológico (CNPq), grants no 304120/2021-9 (JEGS), for financial support.

\appendix
\section{Bumblebee excitations in curved spacetime}\label{a}

In this appendix, we aim to clarify the action (\ref{action1}) used in our study, particularly focusing on the Lorentz violation sector. Our approach is based on the works \cite{leandro,kos1}.
Let us begin by recalling a well-established finding in the literature concerning Lorentz violation. Theories featuring spontaneous local Lorentz and diffeomorphism violation include massless Nambu-Goldstone modes. These modes emerge as field excitations in the minimum of the symmetry-breaking potential. On the other hand, it is also possible to obtain massive modes. Assuming a smooth quadratic potential such as
\begin{equation}
V = \frac{\lambda}{2}(B_{\mu}B^{\mu}\pm b^2)^2,
\end{equation}
then it is possible to show that excitations above the minimum are allowed, so that an alternative Higgs gravitational mechanism can occur in which massive modes involving the metric appear. Here, $\lambda$ represents a positive self-interaction coupling constant with mass dimension one, while $b^{2}$ is a positive constant with squared mass dimension. The $\pm$ sign indicates whether $b_{\mu}$ is spacelike or timelike. Furthermore, the vacuum condition $V=0$ leads to the existence of a vacuum expectation value $<B_{\mu}>=b_{\mu}$ in the following form:
\begin{equation} \label{norma}
g^{\mu\nu}b_{\mu}b_{\nu} = \mp b^{2}.
\end{equation}

To verify this assertion, we proceed by considering the following decomposition:
\begin{equation}
    B_{\mu} = b_{\mu} + \chi_{\mu}.
\end{equation}
Since the VEV defines a preferred direction in spacetime, we can decompose $\chi_{\mu}$ into transverse $A_{\mu}$ and longitudinal $\beta$ modes with respect to $b_{\mu}$ \cite{kos1} \begin{equation} \label{decomp}
\chi _{\mu} = \Tilde{A}_{\mu} + \beta \hat{b}_{\mu},
\end{equation}
where by defining the projection operators $P^{||}_{\mu\nu} = \frac{b_{\mu} b_{\nu}}{b^{\alpha}b_{\alpha}}$ and $P^{\perp}_{\mu\nu} = g_{\mu\nu} - \frac{b_{\mu} b_{\nu}}{b^{\alpha}b_{\alpha}}$,
we have $\Tilde{A}_{\mu} = P^{\perp}_{\mu\nu} \chi^{\nu}$ and $\beta\hat{b}_{\mu} = P^{||}_{\mu\nu} \chi^{\nu}$.
As result, we have to $\Tilde{A}_{\mu}b^{\mu}\approx0$ and $\hat{b}_{\mu}\hat{b}^{\mu} = \mp1$, where $\hat{b}_{\mu} = \frac{b_{\mu}}{\sqrt{b^{2}}}$.  Using the decomposition (\ref{decomp}), the smooth quadratic potential term is given at first order by
\begin{equation}\label{p}
V \approx 4 \lambda[(\hat{b} ^{\alpha}b_{\alpha})\beta]^{2},
\end{equation}
i.e., $V(X)\neq0$, therefore the $\beta$ is a massive mode.

In the literature, it has been demonstrated that these modes experience strong coupling when considering a general spacetime. However, there are two situations in which this difficulty can be overcome. The first scenario arises when we assume that these fluctuations "exist" in a flat spacetime (see \cite{kos1}). However, in this case, it has been shown that the massive mode becomes non-dynamic. The second solution involves selecting a specific spacetime configuration with a preferred direction for the vacuum expectation value (VEV). This choice of preferred direction allows for a viable alternative where the issues of strong coupling can be addressed. Let's explore this scenario in more detail.

In order to address the strong coupling issue arising from the curvature and coupling between longitudinal and transverse modes, let's consider the propagation of bumblebee fluctuations on a symmetric spacetime described by metric (\ref{metricgeral}). We assume a spacelike VEV with only one nonvanishing component, given as: $b_{\mu} = (0, b_r, 0, 0)$. We can determine the radial component using Eq. (\ref{norma}), leading to the following expression:
\begin{equation} \label{vev4}
b_{r} = \frac{b \ell}{r \sqrt{f(r)}}.
\end{equation}
It's worth noting that the VEV choice specified in \eqref{vev4} results in a vanishing field strength, meaning that $b_{\mu\nu}=0$. Additionally, due to the decomposition, this choice leads to $\Tilde{A}_r=0$.

Returning to the action (\ref{action}), we can now replace the decomposition (\ref{decomp}) into it. Let's begin by dealing with the kinetic term of the bumblebee field. It is straightforward to see, albeit a bit tedious, that
\begin{align}
 B_{\mu\nu}B^{\mu\nu}=\chi_{\mu\nu}\chi^{\mu\nu} = \Tilde{F}_{\mu\nu}\Tilde{F}^{\mu\nu} + 4\Tilde{F}^{\mu\nu}(\partial_{\mu}\beta)\hat{b}_{\nu} + 2 (\partial_{\mu}\beta)(\partial^{\mu}\beta) - 2 (\partial_{\mu}\beta)(\partial^{\nu}\beta)(\hat{b}_{\nu}\hat{b}^{\mu}),
\end{align}
where we use that $b_{\mu\nu}=\hat{b}_{\mu\nu}=0$. Note that after the second equality, we recognize that the first term corresponds to the Maxwell term. Moreover, the second term vanishes if we make the assumption that the fields depend solely on the radial coordinate. Furthermore, upon utilizing Eq. (\ref{vev4}), we observe that the usual kinetic term of the longitudinal mode cancels out with another term. Consequently, for this specific choice of the preferred direction in spacetime, the bumblebee kinetic term simplifies to just the Maxwell-type term, which is described by the transverse mode.  In this specific configuration, the massive mode not only becomes decoupled from the non-massive mode but also loses its dynamics. As a consequence, we can describe the massive mode as being "non-dynamic." In this context, we denote the constant associated with the non-dynamic massive mode as $\beta_0$. Indeed, taking all the considerations about the kinetic term and the potential into account (\ref{p}), we derive the action (\ref{action1}).

%\nocite{*} % to test all bib entrys
%\bibliographystyle{unsrt}
%\bibliography{output} % file mwe.bib

%%%%%%%%%%%%%%%%%%%%%%%%%%%%%%%%%%%%%%%%%%%%%%%%%%%%%%%%%%%%%%%%%%%%%%%%%%%%%%%%%%%%%%%%%%%%%%%%%%%%%%%%%%%%%%%%%%%%%%%%%%%5

\end{document}